\documentclass[journal]{new-aiaa}
\usepackage[utf8]{inputenc}
\usepackage{textcomp}
\usepackage{graphicx}
\usepackage{amsmath}
\usepackage[version=4]{mhchem}
\usepackage{siunitx}
\usepackage{longtable,tabularx}
\usepackage{float}
\usepackage{multirow}
\usepackage{comment}
\usepackage[ruled,vlined]{algorithm2e}
\usepackage{mathtools}
\usepackage{autobreak}
\usepackage{xcolor}
\usepackage{lipsum}

\newcommand\blfootnote[1]{%
  \begingroup
  \renewcommand\thefootnote{}\footnote{#1}%
  \addtocounter{footnote}{-1}%
  \endgroup
}

\DeclarePairedDelimiter\norm{\lVert}{\rVert}
\newtheorem{theorem}{Theorem}
\newtheorem{lemma}[theorem]{Lemma}
\newtheorem{prop}{Proposition}

\title{Nonlinear Model Predictive Detumbling of Small Satellites with a Single-axis Magnetorquer\blfootnote{Related work, "Model Predictive Approach for Detumbling an Underactuated Satellite," presented as AIAA 2020-1433 at AIAA Scitech 2020 Forum, Orlando, FL, 6-10 January 2020}}

\author{Kota Kondo \footnote{Student, Department of Mechanical and Aerospace Engineering; kongon1004@gmail.com, AIAA Student Member }}\affil{Kyushu University, Fukuoka-shi, Fukuoka, 819-0395, Japan} 
\author{Ilya Kolmanovsky \footnote{Professor, Department of Aerospace Engineering; ilya@umich.edu, AIAA Associate Fellow.}}\affil{University of Michigan, Ann Arbor, Michigan 48109}

\author{Yasuhiro Yoshimura\footnote{Assistant Professor, Department of Aeronautics and Astronautics.}}\affil{Kyushu University, Fukuoka-shi, Fukuoka, 819-0395, Japan }

\author{Mai Bando\footnote{Associate Professor, Department of Aeronautics and Astronautics.}}
\affil{Kyushu University, Fukuoka-shi, Fukuoka, 819-0395, Japan }

\author{Shuji Nagasaki\footnote{Assistant Professor, Department of Aeronautics and Astronautics.}}
\affil{Kyushu University, Fukuoka-shi, Fukuoka, 819-0395, Japan }

\author{Toshiya Hanada\footnote{Professor, Department of Aeronautics and Astronautics.}}
\affil{Kyushu University, Fukuoka-shi, Fukuoka, 819-0395, Japan}

\begin{document}

\maketitle

\section*{Nomenclature}

{\renewcommand\arraystretch{1.0} 
\noindent\begin{longtable*}{@{}l @{\quad=\quad} l@{}}

$\boldsymbol{B}$, $\boldsymbol{B_0}$   & Earth's magnetic field vector in body and orbital frame, respectively\\ 
$H$ & Hamiltonian \\
$i$ & inclination \\
$J_\text{cost}$  & cost function\\
$\boldsymbol{J}$  & moment of inertia\\
$\boldsymbol{m}$ & magnetic dipole moment\\
$m_\text{max}$   & maximum control input of magnetic torquer \\
$m_x$ &  magnetic moment of the single magnetic torquer along the $x$-axis\\
$N$ & discretized step number on prediction horizon\\
$\boldsymbol{Q}, R_{1}, R_{2}$  & weight matrices \\
$\boldsymbol{Q_{\rm t}}$ & terminal cost \\
$r$ & distance between the satellite and the center of the Earth\\
$\boldsymbol{T}$  & control torque vector  \\
$T_s$   & prediction horizon \\
$\boldsymbol{U}$  & optimal input matrix \\
$V$  & Lyapunov function\\
$v$ & dummy input\\
$\lambda$ & Lagrange multiplier\\
$\boldsymbol{\omega}$ & angular velocity vector\\
$\omega_e$ & argument of perigee\\
$\mu$ & Lagrange multiplier for equality constraint\\
$\theta$ & true anomaly\\
\multicolumn{2}{@{}l}{Subscripts}\\
$i$ & $i$-th time step on prediction horizon \\
$*$ & conditions on prediction horizon\\

\end{longtable*}}

\section{Introduction}
\lettrine{V}{arious} actuators are used in spacecraft to achieve attitude stabilization, including thrusters, momentum wheels, and control moment gyros. \cite{jin2006, fan2002, momemt_gyro}. Small satellites, however, have stringent size, weight, and cost constraints, which makes many actuator choices prohibitive. Consequently, magnetic torquers have commonly been applied to spacecraft to attenuate angular rates~\cite{magnetic_1, magnetic_2}. Approaches for dealing with under-actuation due to magnetic control torques dependency on the magnetic field and required high magnetic flux densities have been previously considered in~\cite{magnetic_3, magnetic_4}.

Generally speaking, control of a satellite that becomes under-actuated as a result of on-board failures has been a recurrent theme in the literature, see e.g., \cite{under-actuated, under-actuated_2} and references therein. Methods for controlling spacecraft with fewer actuators than degrees of freedom are increasingly in demand due to the increased number of small satellite launches \cite{small_sat}.

Magnetic torquers have been extensively investigated for momentum management of spacecraft with momentum wheels \cite{momentum_wheel} and for nutation damping of spin satellites \cite{nutation_damping}, momentum-biased \cite{momemtum_biased}, and dual-spin satellites~\cite{dual_spin}. Nonetheless, severely under-actuated small spacecraft that carry only a single-axis magnetic torquer have not been previously treated.

This note considers the detumbling of a small spacecraft using only a single-axis magnetic torquer. Even with a three-axis magnetic torquer, the spacecraft is under-actuated, while, in the case of only a single axis magnetic torquer, the problem is considerably more demanding. Our note examines the feasibility of spacecraft attitude control with a single-axis magnetic torquer and possible control methods that can be used.

Our specific contributions are as follows. We demonstrate, through analysis and simulations, that the conventional B-dot algorithm for spacecraft detumbling fails with a single-axis magnetic torquer. Also, there has not been any previous analysis of a satellite’s controllability and stabilizability with a single magnetic actuator. We discuss these properties; this discussion motivates consideration of more advanced control approaches such as Nonlinear Model Predictive Control (NMPC) \cite{NMPC, NMPC_2}. Closed-loop simulation results with NMPC are reported, which illustrate the potential of NMPC to perform spacecraft detumbling with a single-axis magnetic torquer. These developments, which show the improved capability to detumble the spacecraft with NMPC as compared to the classical B-dot law in the case of a single magnetic torquer, contribute to advancements in small satellite technology.


\section{Spacecraft Rotational Dynamics}

The body-fixed frame of a rigid spacecraft is assumed to be located at the center of mass and to be aligned with the principal axes of inertia. The evolution of body frame components of spacecraft angular velocity vector is described by the classical Euler's equations \cite{dynamics}:
\begin{equation}\label{dynamics}
    \boldsymbol{J} \boldsymbol{\dot\omega}+\boldsymbol{\omega} \times \boldsymbol{J} \boldsymbol{\omega}=\boldsymbol{T}
\end{equation}

A single-axis magnetic torquer interacts with the Earth’s local magnetic field and generates control torque according to~\cite{torque}:
\begin{equation}\label{torque_in_matrix}
    \boldsymbol{T}=\boldsymbol{m} \times \boldsymbol{B}
\end{equation}

Given a single-axis magnetic torquer with the coil along the $x$-axis, the control torque with the single-axis magnetorquer is written as $\boldsymbol{T}=[0, \ -B_z m_x, \ B_y m_x]^T$. Thus, no torque is generated about the $x$-axis, which makes angular rate stabilization challenging. By aggregating the above equations, we obtain
\begin{equation}\label{dynamics_combined}
    \begin{bmatrix}
        \dot\omega_{x}\\
        \dot\omega_{y}\\
        \dot\omega_{z}\\
    \end{bmatrix}
    =
    \begin{bmatrix}
        \frac{1}{J_x}{(J_y-J_z)\omega_y\omega_z}\\
        \frac{1}{J_y}{(J_z-J_x)\omega_z\omega_x-\frac{B_z m_x}{J_{y}}}\\
        \frac{1}{J_z}{(J_x-J_y)\omega_x\omega_y+\frac{B_y m_x}{J_{z}}}\\
    \end{bmatrix}
\end{equation}
where $m_x$ is the control input.

\section{Detumbling Control Law}

\subsection{B-dot Algorithm} \label{B-dot_algorithm}

The conventional approach to detumbling small satellites with magnetic torquers exploits the B-dot algorithm~\cite{B-dot}. The B-dot algorithm’s principle is to add damping through control moments, which leads to a reduction in spacecraft angular velocities. This section analyzes closed-loop stability with the B-dot law in the case of a single-axis magnetic torquer. Considering the Euler's equations for spacecraft dynamics given is Eq.~\eqref{dynamics}, define a Lyapunov function~\cite{B-dot} as
\begin{equation}\label{Lyapunov_function}
    V(\boldsymbol{\omega})
    =
    \frac{1}{2} \boldsymbol{\omega}^T \boldsymbol{J} \boldsymbol{\omega}
\end{equation}

This Lyapunov function is positive everywhere except when $ \boldsymbol{\omega} = 0$  ( i.e., the equilibrium point). With the use of Eq.~\eqref{dynamics}, its time derivative along trajectories of the system is found as 
\begin{eqnarray}\label{dot_Lyapnov_function}
    \dot V(\boldsymbol{\omega}) 
    & = & \boldsymbol{\omega}^T \boldsymbol{J} \boldsymbol{\dot\omega} \\
    & = & \boldsymbol{\omega}^T (-\boldsymbol{\omega} \times \boldsymbol{J} \boldsymbol{\omega}+\boldsymbol{T}) \\
    & = & \boldsymbol{\omega}^T \boldsymbol{T}
\end{eqnarray}

The conventional B-dot feedback law in \cite{B-dot_2, B-dot_3} generates each axis magnetic dipole moment as follows. 
For instance, for the $x$-axis,
\begin{equation}\label{B-dot_input}
    m_x = -m_\text{max} \frac{\dot B_x}{||\dot B_x||}
\end{equation}
where $m_{\rm max}$ is the maximum magnitude of the magnetic dipole moment.

The time derivative of Earth’s magnetic field vector $\boldsymbol{B}$ with respect to an inertial frame is given by
\begin{equation}\label{magnetic_field_wrt_I}
    ^I\boldsymbol{\dot B} = \boldsymbol{\dot B} + \boldsymbol{\omega} \times \boldsymbol{B}
\end{equation}
where the left superscript $I$ on $\boldsymbol{B}$ indicates ''with respect to an inertial frame.'' Assuming sufficiently large angular velocity, $\boldsymbol{\omega}$, so that $^I\boldsymbol{\dot B}$ is small enough in magnitude compared to $\boldsymbol{\dot B}$, where the latter is the derivative of Earth’s magnetic field vector with respect to a body fixed frame, Eq.~\eqref{magnetic_field_wrt_I} can be approximated as \cite{B-dot_3, magnetic_1}.
\begin{eqnarray}\label{magnetic_field_approximation}
    0 
    & \approx & 
    \boldsymbol{\dot B} + \boldsymbol{\omega} \times \boldsymbol{B} \\
    \Rightarrow 
    \boldsymbol{\dot B}
    & \approx &
    -\boldsymbol{\omega} \times \boldsymbol{B}
\end{eqnarray}

Following ~\cite{B-dot_stable_proof}, which treated the case of three single-axis magnetic torquers, suppose we proceed with closed-loop stabilizability analysis by computing the time derivative of $V$ 
along closed-loop system trajectories. In the case of a single-axis magnetic actuation, we obtain
\begin{eqnarray}\label{dot_V_kind_of_long}
    \dot V(\boldsymbol{\omega}) 
    &=& -(\boldsymbol{\omega} \times \boldsymbol{B})^T \boldsymbol{m}\\
    &=& -\frac{m_\text{max}}{||\boldsymbol{\dot B}||}(\boldsymbol{\omega} \times \boldsymbol{B})^T [\omega_y B_z-\omega_z B_y, 0, 0]^{T}\\
    &=& -\frac{m_\text{max}}{||\boldsymbol{\dot B}||}(\omega_y B_z-\omega_z B_y)^2
\end{eqnarray}

It is clear that, although $\dot V(\boldsymbol{\omega}) \leq 0$, the expression for $\dot{V}$ does not depend on $\omega_x$. Furthermore, since $V(t)=V(\boldsymbol{\omega}(t))$ is a non-increasing function of $t$, $\boldsymbol{\omega}$ is bounded. Moreover $\ddot{V}$ is a continuous function of $\boldsymbol{B}$ and  $\boldsymbol{\omega}$, which are bounded. Hence, $\dot V(t)$ is uniformly continuous in time~\cite{Barbalet's_lemma}. Therefore, by Barbalat's lemma, we conclude $\lim_{t\to\infty} \dot V(t)=0$, which indicates that in the limit as $t\to\infty$ either $\omega_y B_z=\omega_z B_y=0$ or $\omega_y=\omega_z=0$. In the either case, since $\omega_x$ can be arbitrary, B-dot algorithm appears to be incapable of detumbling a satellite with a single-axis magnetic actuation. This is confirmed by our subsequent numerical simulations.

\subsection{Controllability Analysis}

This section discusses the controllability properties of the satellite angular velocity dynamics with the single-axis magnetic actuation.  
Note that both three and two-axis magnetically actuated satellites have been shown to be controllable ~\cite{Controllability_2,Finite_time}.  The single-axis magnetic actuation case is more challenging since there is only a single control input; this case has not previously been addressed in the literature.

For the case of single-axis magnetic actuation,  local weak controllability in the sense of \cite{R_Hermann_Nonlinear_controllability_and_observability,cheng_analysis_of_non_linear_ctrl_sys_for_local_weak_ctrlbility} can be demonstrated. The weak local controllability is necessary for local controllability. It implies that the set of reachable states at a given time from a given state starting at another given time instant contains an open neighborhood of the state space. Clearly, weak local controllability is necessary but not sufficient for local controllability.


Note that equations of motion (\ref{dynamics_combined}) can be written as
\begin{eqnarray}\label{dynamics_for_controllability_proof}
    \boldsymbol{\dot\omega}
    &=&
    f_{0}(\boldsymbol{\omega},t)+f_{1}(\boldsymbol{\omega},t)\ m_x\\
    &=&
    \begin{bmatrix}
        \frac{1}{J_x}(J_y-J_z)\omega_y\omega_z\\
        \frac{1}{J_y}(J_z-J_x)\omega_z\omega_x\\
        \frac{1}{J_z}(J_x-J_y)\omega_x\omega_y\\
    \end{bmatrix}
    +
    \begin{bmatrix}
        0\\
        -\frac{B_z}{J_{y}}\\
        \frac{B_y}{J_{z}}\\
    \end{bmatrix}
    m_x.
\end{eqnarray}
The system is time-varying as $B_y$ and $B_z$ depend on the spacecraft position in orbit and hence on time.
The necessary and sufficient conditions for local weak controllability of a time-varying nonlinear system can be obtained by extending the state vector with the time, $t$, as an extra state with the dynamics $\dot{t}=1$, and analyzing the resulting
autonomous system. This leads to conditions such as  Theorem~4 in~\cite{Martinelli_Rank_Conditions_for_Observability_and_Controllability_for_Time-varying_Nonlinear_Systems} which we adopt here.

Define the operators $\langle \xi,\eta \rangle$ and $[\xi,\eta]$ for two time-varying vector fields $\xi$ an $\eta$ as
\begin{equation}\label{symmetric product}
        \langle \xi,\eta \rangle 
        \equiv
        [\xi,\eta]-\frac{\partial \xi}{\partial t},
\end{equation}
\begin{equation}\label{Lie_bracket_equation_2}
    [\xi,\eta]=\frac{\partial \eta}{\partial \omega}\xi-\frac{\partial \xi}{\partial \omega}\eta.
\end{equation}
Note that $[\xi,\eta]$ is the conventional Lie Bracket.
The controllability distribution $\triangle$ can now be defined for  our time-varying nonlinear system  based on Algorithm \ref{controllability_distribution_algorithm}.

\begin{algorithm}[H]\label{controllability_distribution_algorithm}
\SetAlgoLined
 Set $\triangle_{0}={\rm span}\{f_{1}\}$ and $k=0$\;
 
 \While{{$\rm dim$}$(\triangle_{k} \oplus \langle \triangle_{k}, f_{0} \rangle \oplus [\triangle_{k}, f_{1}])$ > {$\rm dim$}$(\triangle_{k})$}{
  Set $\triangle_{k+1}=\triangle_{k}\oplus \langle \triangle_{k}, f_{0} \rangle \oplus [\triangle_{k}, f_{1}]$\;
  $k=k+1$\;
 }
 \caption{Controllability distribution for time-variant nonlinear systems}
\end{algorithm}
$\oplus$ sums up the span of two vector fields.

The necessary and sufficient conditions for local weak controllability in~\cite{Martinelli_Rank_Conditions_for_Observability_and_Controllability_for_Time-varying_Nonlinear_Systems} lead to the following result: Algorithm~\ref{controllability_distribution_algorithm} converges in at most $2$ steps for the system~(\ref{dynamics_for_controllability_proof}) and $\triangle_{2}$ is nonsingular and has rank $3$  at a given $\boldsymbol{\omega}_0$ and $t_0$ if and only if the system is locally weakly controllable from ($\boldsymbol{\omega}_0$, $t_0$).

By examining the form of $\triangle_{2}$ (see calculations in the Appendix), we observe that $J_y \neq J_z$ (unequal moments of inertia about the two principal axes which are orthogonal to the axis along which the magnetic actuator is aligned) is a necessary condition for weak local controllability (and hence for local controllability).  This is also apparent from (\ref{dynamics_for_controllability_proof}) as the angular velocity component about $x$-axis becomes decoupled from the rest of the dynamics if $J_y=J_z$.

Furthermore, the numerical evaluation of $\triangle_{2}$ along the orbits used in our NMPC simulations with IGRF model (\ref{dipole_model}) for $B_y$ and $B_z$
confirms that the rank of $\triangle_{2}$ is equal to $3$ and hence the system is locally weakly controllable.

We note that we cannot establish a stronger property of (small-time) local controllability of the satellite with a single-axis magnetic actuation with the above analysis. While it appears to hold in our numerical simulations, such property is much harder to demonstrate and is left to future research. Nevertheless, conditions for local weak controllability are necessary for local controllability and hence are useful.

\subsection{Stabilizability Analysis}

Existing approaches \cite{Pseudoinverse,dynamics,lovera2005global,Controllability_2} to stabilizing the spacecraft with magnetic actuation typically rely on the application of the theory of averaging \cite{Khail_Nonlinear_systems}. Following this route in the case of single-axis magnetic actuation encounters technical difficulties as we now illustrate.

Consider the equations of motion given by Eq.~\eqref{dynamics_for_controllability_proof} and 
suppose a control law of the form, 
\begin{equation}\label{equ:ave_control_law}
m_x = \epsilon^2 \bar{m}_x(
\frac{\boldsymbol{\omega}}{\epsilon},t)
\end{equation} has been specified, where $\epsilon>0$ is a small parameter.  Let $$\boldsymbol{h}=\frac{\boldsymbol{\omega}}{\epsilon}=\left[ \begin{array}{ccc} h_x & h_y & h_z \end{array} \right]^{\rm T},$$
then 
\begin{eqnarray}\label{equ:ave_closed_loop}
    \dot{h}_x & = &    \epsilon  \frac{1}{J_x}(J_y-J_z)h_y h_z \nonumber \\
    \dot{h}_y &=&    \epsilon \frac{1}{J_y}(J_z-J_x)h_z h_x-\frac{\epsilon}{J_{y}} B_z(t)\bar{m}_x(\boldsymbol{h},t)
        \\
    \dot{h}_z & = &    \epsilon \frac{1}{J_z}(J_x-J_y)h_xh_y
        +\frac{\epsilon}{J_{z}} B_y(t)\bar{m}_x(\boldsymbol{h},t) \nonumber
\end{eqnarray}
is in the form to which the theory of averaging~\cite{Khail_Nonlinear_systems} can be applied.
Letting
\begin{eqnarray}\label{equ:inputs_ave_closed_loop}
    u_y(\boldsymbol{h})=-\frac{\epsilon}{J_y}\lim_{T \to \infty} \frac{1}{T}\int_0^T B_z(t) \bar{m}_x(\boldsymbol{h},t) dt \nonumber
        \\
    u_z(\boldsymbol{h})=\frac{\epsilon}{J_z}\lim_{T \to \infty} \frac{1}{T} \int_0^T B_y(t) \bar{m}_x(\boldsymbol{h},t) dt   \nonumber
\end{eqnarray}
and assuming that the limits exist,
the averaged version of Eq.~\eqref{equ:ave_closed_loop} has the form,
\begin{eqnarray}\label{equ:ave_closed_loop_ave}
    \dot{\bar{h}}_x &=&  \epsilon \frac{1}{J_x}(J_y-J_z)\bar{h}_y \bar{h}_z \nonumber \\
    \dot{\bar{h}}_y &=&  \epsilon \frac{1}{J_y}(J_z-J_x)\bar{h}_z \bar{h}_x+u_y(\bar{\boldsymbol{h}})\\
    \dot{\bar{h}}_z &=&  \epsilon \frac{1}{J_z}(J_x-J_y)\bar{h}_x \bar{h}_y
        +u_z(\bar{\boldsymbol{h}}) \nonumber
\end{eqnarray}
By the straightforward application of Brockett's necessary condition~\cite{Brockett}, the averaged system given by Eq.~\eqref{equ:ave_closed_loop_ave} is not  smoothly or even continuously stabilizable, i.e., a stabilizing time-invariant control law given by $u_y(\bar{\boldsymbol{h}})$,
$u_z(\bar{\boldsymbol{h}})$, if exists, would have to be discontinuous.  This conclusion is consistent with the interpretation of the averaged system dynamics as similar to  a rigid body controlled by two control torques; such a system is known to be not smoothly or even continuously stabilizable by time-invariant feedback laws. Consequently,  $\bar{m}_x(
\frac{\boldsymbol{\omega}}{\epsilon},t)$
would have to be discontinuous as a function of $\boldsymbol{\omega}$.  Unfortunately, the theory of averaging  \cite{Khail_Nonlinear_systems, 1984spacecraft} assumes smoothness (twice continuous differentiability) in \cite{Khail_Nonlinear_systems} of the right hand side of ordinary differential equations being averaged.
 
Hence, there is a complication in using the classical averaging theory to develop conventional control laws for the case of single-axis magnetic actuation. On the other hand, NMPC, if suitably formulated, is able to stabilize systems that are not smoothly or even continuously stabilizable, including underactuated spacecraft \cite{petersen2017model}.

\subsection{NMPC Formulation}

This section formulates an NMPC approach to detumbling the satellite with the nonlinear dynamics represented by Eq.\eqref{dynamics_combined} based on the following receding horizon optimal control problem,
\begin{equation}\label{NMPC_formulation}
    \centering
    \begin{array}{rl}
    {\rm minimize} & J_\text{cost} =\frac{1}{2}\boldsymbol{\omega}^{\rm T} (t+T_s)\boldsymbol{Q}_{\rm t}\boldsymbol{\omega}(t+T_s)+
    \int^{t+T_s}_t \{\frac{1}{2}
    (\boldsymbol{\omega}^{\rm T}(\tau) \boldsymbol{Q} \boldsymbol{\omega}(\tau)+R_1 m_x^{2}(\tau))-R_2 v(\tau)\}d\tau\\
    
    {\rm subject \ to} & \boldsymbol{J} \boldsymbol{\dot\omega}+
    \boldsymbol{\omega} \times \boldsymbol{J} \boldsymbol{\omega}=\boldsymbol{T}\\
    
    \ & \boldsymbol{T}=\boldsymbol{m} \times \boldsymbol{B}, \ \ {\rm where}\ \  \boldsymbol{m} = [m_x, 0, 0]^{\rm T}\\
    
    \ & m_x^2+v^2-m_\text{max}^2=0\\
    \end{array}
\end{equation}
where $t$ is the current time instant, $T_s$ is the prediction horizon, $\boldsymbol{Q}$, $\boldsymbol{Q_{\rm t}}$, $R_1$, and $R_2$ are positive-definite weight matrices, and $\boldsymbol{Q_{\rm t}}$ is terminal cost.
The auxiliary input, $v$, is introduced following \cite{NMPC_formulation_1} to enforce the control constraints by recasting them as equality constraints in Eq.~\eqref{NMPC_formulation}. The negative sign preceding $R_2$ in the cost function being minimized promotes keeping $v$ positive and control constraints strictly satisfied. This receding horizon optimal control problem is chosen as it is synergistic with the continuation/generalized minimal residual method (C/GMRES) method \cite{Seguchi_Nonlinear_receding_horizon_control_of_an_underactuated_hovercraft}. Reference \cite{huang2015nonlinear} provides a comparison of different strategies to handle inequality constraints in such a setting.


\subsection{NMPC with C/GMRES Algorithm}

The C/GMRES method \cite{Seguchi_Nonlinear_receding_horizon_control_of_an_underactuated_hovercraft} is applied to design NMPC based on Eq.\eqref{NMPC_formulation}.  As C/GMRES method has small computational footprint, its use is advantageous in small satellites with limited computational and electric power. Following C/GMRES method, the problem is first discretized as follows: 
\begin{equation}\label{Real_time_Update_1}
    \boldsymbol{\omega}^*_{i+1}(t)=\boldsymbol{\omega}^*_{i}(t)+f(\boldsymbol{\omega}^*_{i}(t),\boldsymbol{u}^*_{i}(t))\Delta\tau
\end{equation}
\begin{equation}\label{Real_time_Update_2}
    \boldsymbol{\omega}^*_0(t)=\boldsymbol{\omega}(t)
\end{equation}
\begin{equation}\label{Real_time_Update_3}
    C(\boldsymbol{\omega}^*_{i}(t),\boldsymbol{u}^*_{i}(t))=0
\end{equation}
\begin{equation}
    J_\text{cost}=\psi(\boldsymbol{\omega}^*_N(t))+\sum^{N-1}_{i=0}L(\boldsymbol{\omega}^*_{i}(t),\boldsymbol{u}^*_{i}(t))\Delta\tau
\end{equation}
 where  
$ f(\boldsymbol{\omega},\boldsymbol{u})$  is the right hand side of equations of motion in Eq.\eqref{NMPC_formulation}, $C(\boldsymbol{\omega},\boldsymbol{u})$ 
is the equality constraint in Eq.\eqref{NMPC_formulation},
$L(\boldsymbol{\omega},\boldsymbol{u}) = \frac{1}{2}(\boldsymbol{\omega}^T \boldsymbol{Q} \boldsymbol{\omega}+R_1 m_x^2 )-R_2 v,$ and $\Delta\tau=T_s/N$.
Setting the initial state of the discretized problem to the current angular velocity vector as $\boldsymbol{\omega}^*_0(t)=\boldsymbol{\omega}(t)$, 
a sequence of control inputs $\{\boldsymbol{u}^*_i(t)\}^{N-1}_{i=0}$ is found at each time instant $t$; then the control given to the system is based on the first element of this sequence and is defined as $\boldsymbol{u}(t)=\boldsymbol{u}^*_0(t)$.

The solution of the discretized problem is based on introducing the
Hamiltonian, $H$, as
\begin{equation}
    H(\boldsymbol{\omega},\boldsymbol{\lambda},\boldsymbol{u},\boldsymbol{\mu})=L(\boldsymbol{\omega},\boldsymbol{u})+\boldsymbol{\lambda}^T f(\boldsymbol{\omega},\boldsymbol{u})+\boldsymbol{\mu}^T C(\boldsymbol{\omega},\boldsymbol{u})
\end{equation}
where $\boldsymbol{\lambda}$ is the vector of co-states and $\boldsymbol{\mu}$ is the Lagrange multiplier associated with the equality constraint. The first-order necessary conditions for optimality dictate \cite{necessary_conditions_for_inputs} that  $\{\boldsymbol{u}^*_i(t)\}^{N-1}_{i=0}$,
$\{\boldsymbol{\mu}^*_i(t)\}^{N-1}_{i=0}$,
$\{\boldsymbol{\lambda}^*_i(t)\}^{N-1}_{i=0}$, 
satisfy the following conditions:
\begin{equation}\label{Real_time_Update_4}
    H_{\boldsymbol{u}} (\boldsymbol{\omega}^*_{i}(t),\boldsymbol{\lambda}^*_{i+1}(t),\boldsymbol{u}^*_{i}(t),\boldsymbol{\mu}^*_{i}(t))=0
\end{equation}
\begin{equation}\label{Real_time_Update_5}
    \boldsymbol{\lambda}^*_i(t)
    =
    \boldsymbol{\lambda}^*_{i+1}(t)
    +
    H_{\boldsymbol{\omega}}^T(\boldsymbol{\omega}^*_{i}(t),\boldsymbol{\lambda}^*_{i+1}(t),\boldsymbol{u}^*_{i}(t),\boldsymbol{\mu}^*_{i}(t))\Delta\tau
\end{equation}
\begin{equation}\label{Real_time_Update_6}
    \boldsymbol{\lambda}^*_N(t)
    =
    \psi^T_{\boldsymbol{\omega}}(\boldsymbol{\omega^*_N}(t))
\end{equation}

To determine $\{\boldsymbol{u}^*_i(t)\}^{N-1}_{i=0}$ and $\{\boldsymbol{\mu}^*_i(t)\}^{N-1}_{i=0}$, which satisfy Eqs.(\ref{Real_time_Update_1}--\ref{Real_time_Update_3}) and (\ref{Real_time_Update_4}--\ref{Real_time_Update_6}), we define a vector of the inputs and multipliers in Eq.~\eqref{U_input_formulation} as

\begin{equation}\label{U_input_formulation}
    \boldsymbol{U}(t)=[{m_{x}}_{0}^*(t), \ v_0^*(t), \ \mu_0^*(t), \ \dots, \ {m_{x}}_{N-1}^{*}(t), \ v_{N-1}^*(t), \ \mu_{N-1}^*(t)]^T
\end{equation}

This vector has to satisfy the equation,

\begin{equation}\label{F_formulation}
    \boldsymbol{F}(\boldsymbol{U}(t),\boldsymbol{\omega}(t),t)
    \equiv
    \begin{bmatrix}
    (\frac{\partial H}{\partial u})^T (\boldsymbol{\omega_0^*}(t),u_0^*(t),\lambda_1^*(t),t)\\
    m_{x0}^2+v_0^2-m_\text{max}^2=0\\
    \vdots\\
    (\frac{\partial H}{\partial u})^T (\boldsymbol{\omega_{N-1}^*}(t),u_{N-1}^*(t),\lambda_N^*(t),t+T)\\
    m_{xN-1}^2+v_{N-1}^2-m_\text{max}^2\\
    \end{bmatrix}
    =\textbf{0} \ ,
\end{equation}   
\begin{equation}\label{F_H_formulation}
    \begin{array}{lcl}
        {\rm where}\ \ H &=& \frac{1}{2} (\boldsymbol{\omega}^T \boldsymbol{Q} \boldsymbol{\omega}+R_1 m_x^2 
        )-R_2 v + \lambda_x\{\frac{1}{J_x}{(J_y-J_z)\omega_y\omega_z}\} + \lambda_y\{\frac{1}{J_y}{(J_z-J_x)\omega_z\omega_x-B_z m_x}\} \\
        &+&
        \lambda_z\{\frac{1}{J_z}{(J_x-J_y)\omega_x\omega_y+B_y m_x}\} + \mu\{m_x^2+v^2-m_\text{max}^2\}
    \end{array}
\end{equation}

In C/GMRES \cite{Seguchi_Nonlinear_receding_horizon_control_of_an_underactuated_hovercraft, Ohtsuka_2}, Eq.~\eqref{F_formulation}, which has to haold at each time instant, $t$, is replaced by a stabilized version, 
\begin{equation}\label{dot_F_formulation}
    \frac{d}{dt}\boldsymbol{F}(\boldsymbol{U}(t),\boldsymbol{\omega}(t),t)
    =
    -\zeta\boldsymbol{F}(\boldsymbol{U}(t),\boldsymbol{\omega}(t),t)
    \ \ \ \ \ (\zeta>0)
\end{equation}
and then by
\begin{equation}\label{F_and_U_to_solve}
    \frac{\partial \boldsymbol{F}}{\partial \boldsymbol{U}} \boldsymbol{\dot U}(t) 
    =
    -\zeta\boldsymbol{F}-\frac{\partial \boldsymbol{F}}{\partial \boldsymbol{U}}\boldsymbol{\dot \omega}(t)-\frac{\partial \boldsymbol{F}}{\partial t}
\end{equation}
where arguments are omitted.

Finally, $\dot{U}$ can be determined from Eq.~\eqref{F_and_U_to_solve} with C/GMRES resulting in a form of a predictor-corrector strategy.  


\section{Simulation Results}

This section presents simulation results of both the B-dot algorithm and NMPC for a spacecraft in a sun-synchronous orbit with orbital elements given in Table \ref{Six_elements}.

\subsection{Orbital Elements}

\begin{table}[H]
\caption{\label{Six_elements} Six elements of Aeolus (sun synchronous orbit)}
\centering
\begin{tabular}{ c c }
    \hline
    Semi-major axis & 6691.6 [km]\\
    \hline
    Eccentricity & 0.00046440\\
    \hline
    Inclination & 96.700[deg]\\
    \hline
    Right Ascension of Ascending Node &  100.90 [deg]\\
    \hline
    Argument of perigee &  119.70 [deg]\\
    \hline
    Mean anomaly &  240.49 [deg]\\
    \hline
\end{tabular}
\end{table}


\subsection{Earth's Magnetic Field Model}

\begin{figure}[H]
\centering
\includegraphics[width=0.75\textwidth]{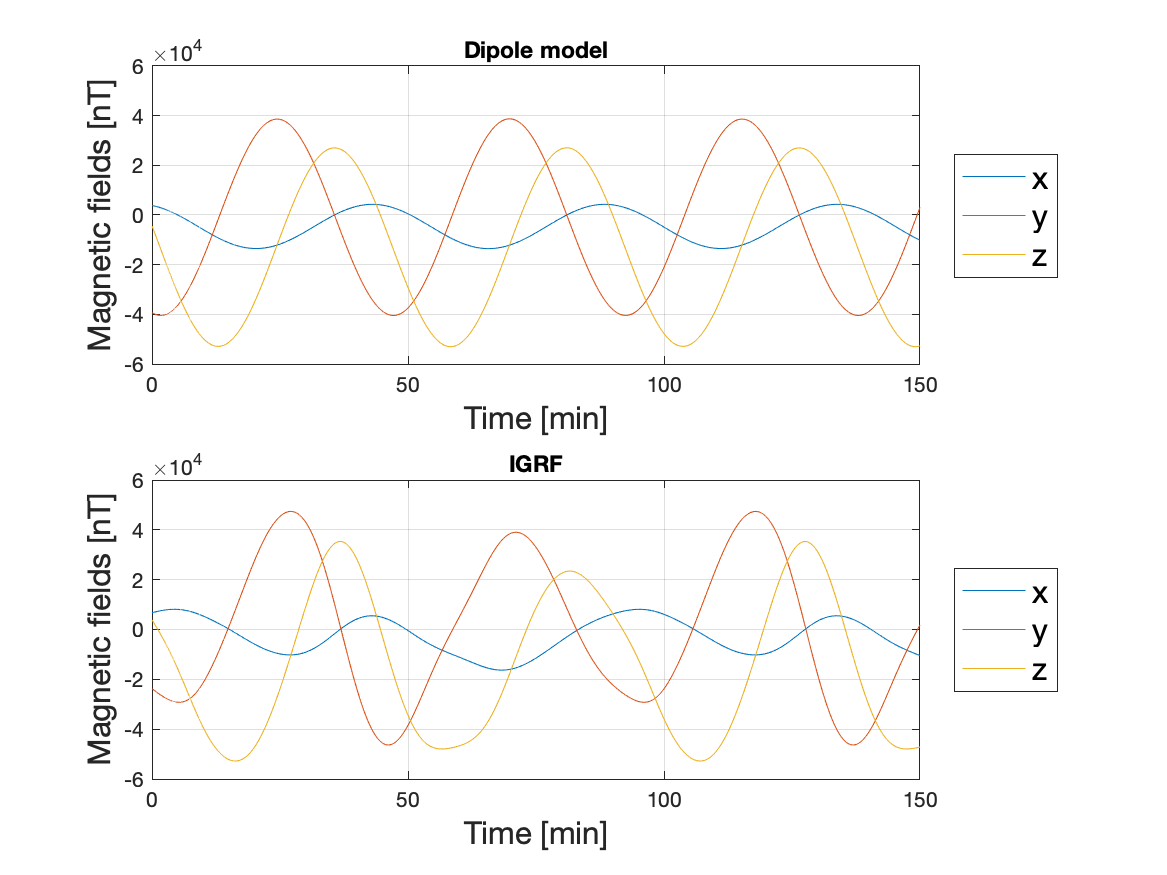}
\caption{Magnetic field in dipole and IGRF on the sun-synchronous orbit.}
\label{fig:magnetic_field_Aeolus}
\end{figure}

We employ International Geomagnetic Reference Field (IGRF) \cite{IGRF} as an Earth's magnetic model in our simulations. However, to reduce the onboard computational complexity, NMPC uses a simpler magnetic dipole model described in Eq.~\eqref{dipole_model} \cite{dipole_model}.

\begin{equation}\label{dipole_model}
\begin{array}{c}
    \begin{bmatrix}
    B_{0x}\\
    B_{0y}\\
    B_{0z}\\
    \end{bmatrix}
    =D_m
    \begin{bmatrix}
    \frac{3}{2}\sin{i}\sin{2\eta}\\
    -\frac{3}{2}\sin{i}\left(\cos{2\eta}-\frac{1}{3}\right)\\
    -\cos{i}\\
    \end{bmatrix}
\end{array}
\end{equation}
where $\eta = \theta+\omega_{e}$, $D_m=-\frac{M_e}{r^3}$, $M_e=8.1\times10^{25}$ [gauss $\cdot$ cm$^{3}$], $r$ is a distance between the satellite and the center of the Earth, $\theta$ is true anomaly, and $\omega_e$ is argument of perigee. Figure \ref{fig:magnetic_field_Aeolus} shows that there is a slight discrepancy between the two models on the sun-synchronous orbit, which, as we will see from the simulation results, will not preclude NMPC controller from achieving detumbling.

\subsection{Comparison of B-dot and NMPC on Asymmetric Satellite}

This section demonstrates NMPC's advantages over the B-dot algorithm, using a general satellite model whose moments of inertia are given in Table \ref{moment_of_inertia}. The maximum magnetic moment is set to 1.0 [A $\cdot$ m$^2$].

\begin{table}[H]
\caption{\label{moment_of_inertia} Moment of inertia of the asymmetric satellite }
\centering
\begin{tabular}{ c c c c }
    \hline
    Moments of inertia & $J_x$ & $J_y$ & $J_z$\\
    \hline
    Value [kg $\cdot$ m$^2$] & 0.020 & 0.030 & 0.040\\ 
    \hline
\end{tabular}
\end{table}


\subsubsection{The B-dot algorithm}

As shown in Eq.\ref{B-dot_input}, the B-dot law finds control inputs depending the Earth's magnetic field. Following the common practice, to avoid the B-dot law generating unnecessary control inputs, its implementation is based on
\begin{equation}\label{B-dot_input_advance}
    m_x = \begin{cases}
      0, & \text{if}\ \dot B_x < 10^{-7} \\
      -m_\text{max} \frac{\dot B_x}{||\dot B_x||}, & \text{otherwise}
    \end{cases}
\end{equation}

\subsubsection{NMPC}

We here demonstrate NMPC's capability of detumbling the spacecraft. Simulations are conducted with the NMPC parameters listed in Table~\ref{table_NMPC_properties_asymmetric}, which were determined by trial and error. 

\begin{table}[H] 
\caption{\label{table_NMPC_properties_asymmetric} NMPC properties for the asymmetric satellite }
\centering
\begin{tabular}{c c c c c c c}
    \hline
    $T_s$ & \textit{$Q$} & \textit{$Q_{\rm t}$} & \textit{$R_1$} & \textit{$R_2$} & $N$ & $\Delta \tau$\\
    \hline
    10 [sec] & diag([$10^{4}$, $10^{2}$, $5 \times 10$])  & diag([$10^{4}$, $10^{2}$, $5 \times 10$]) & $10^{-1}$ & $10^{-1}$ & $10$ & 1.0 [sec] \\
    \hline
\end{tabular}
\end{table}
\noindent
where $T_s$ is prediction horizon, $Q$, $Q_{\rm t}$ $R_1$, and $R_2$ are weight matrices, $N$ is discretized step number along prediction horizon, and $\Delta \tau=T_s/N$.

The initial conditions for the four study cases for which simulation results are reported below are chosen randomly with angular velocity components between -3.0 and 3.0 [deg/s]. The initial angular velocities and the results are all given in Table~\ref{initital_conditions}. These four simulations are representatives of a larger number of simulation case studies that we have performed. 

\begin{table}[H] 
\caption{\label{initital_conditions} Initial conditions and results}
\centering
\begin{tabular}{c c c c c c}
    & $\omega_x$ [deg/s] & $\omega_y$ [deg/s] & $\omega_z$ [deg/s] & B-dot detumbled? & NMPC detumbled?\\
    \hline
    Case 1 & 2.429286 & 2.878490 & -0.366780 & No & Yes\\
    \hline
    Case 2 & -1.576299 & -0.246907 & 2.778531 & No & Yes\\
    \hline
    Case 3 & 0.047150 & -2.486905 & -1.425107 & Yes & Yes\\
    \hline
    Case 4 & -0.626909 & -0.795380 & 2.927892 & No & No\\
    \hline
\end{tabular}
\end{table}

\begin{figure}[H]
\centering
\includegraphics[width=0.75\textwidth]{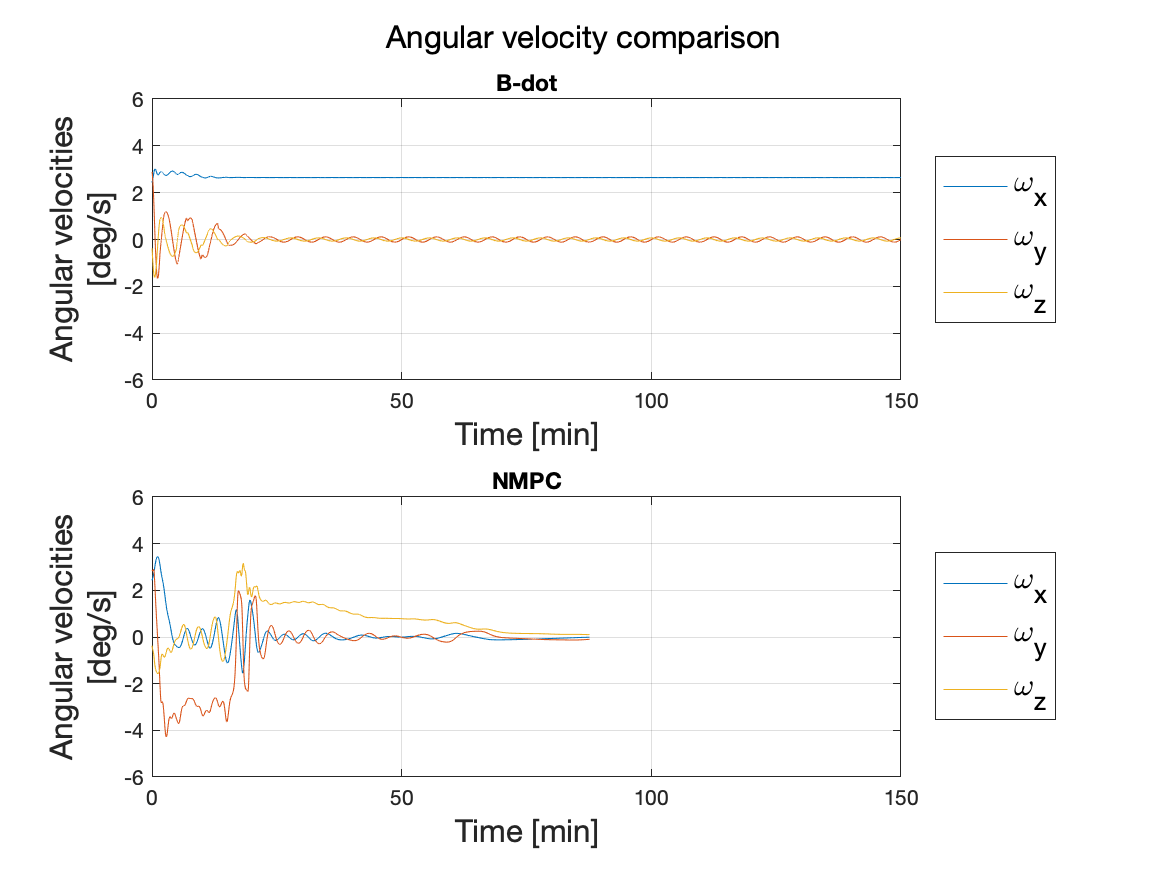}
\caption{Case 1: time history of angular velocities.}
\label{fig:Case1}
\end{figure}

\begin{figure}[H]
\centering
\includegraphics[width=0.75\textwidth]{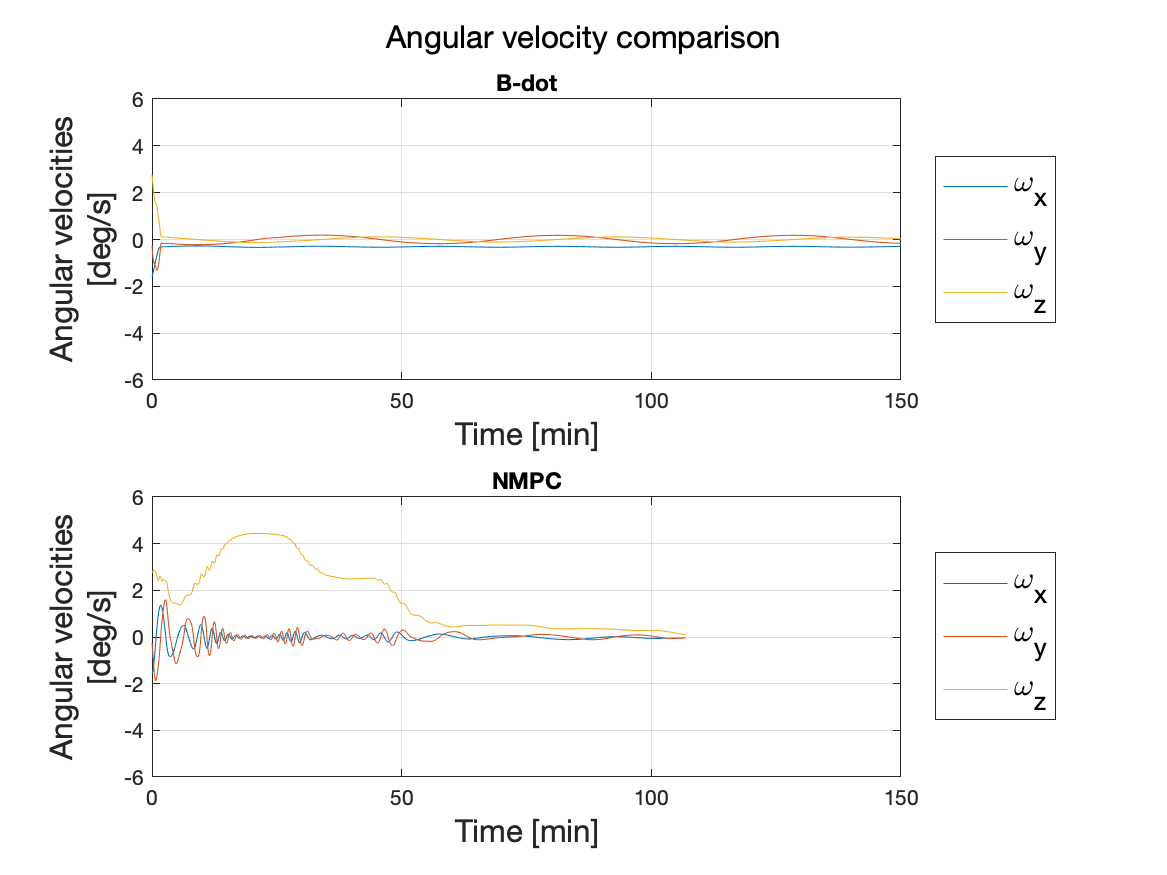}
\caption{Case 2: time history of angular velocities.}
\label{fig:Case2}
\end{figure}

\begin{figure}[H]
\centering
\includegraphics[width=0.75\textwidth]{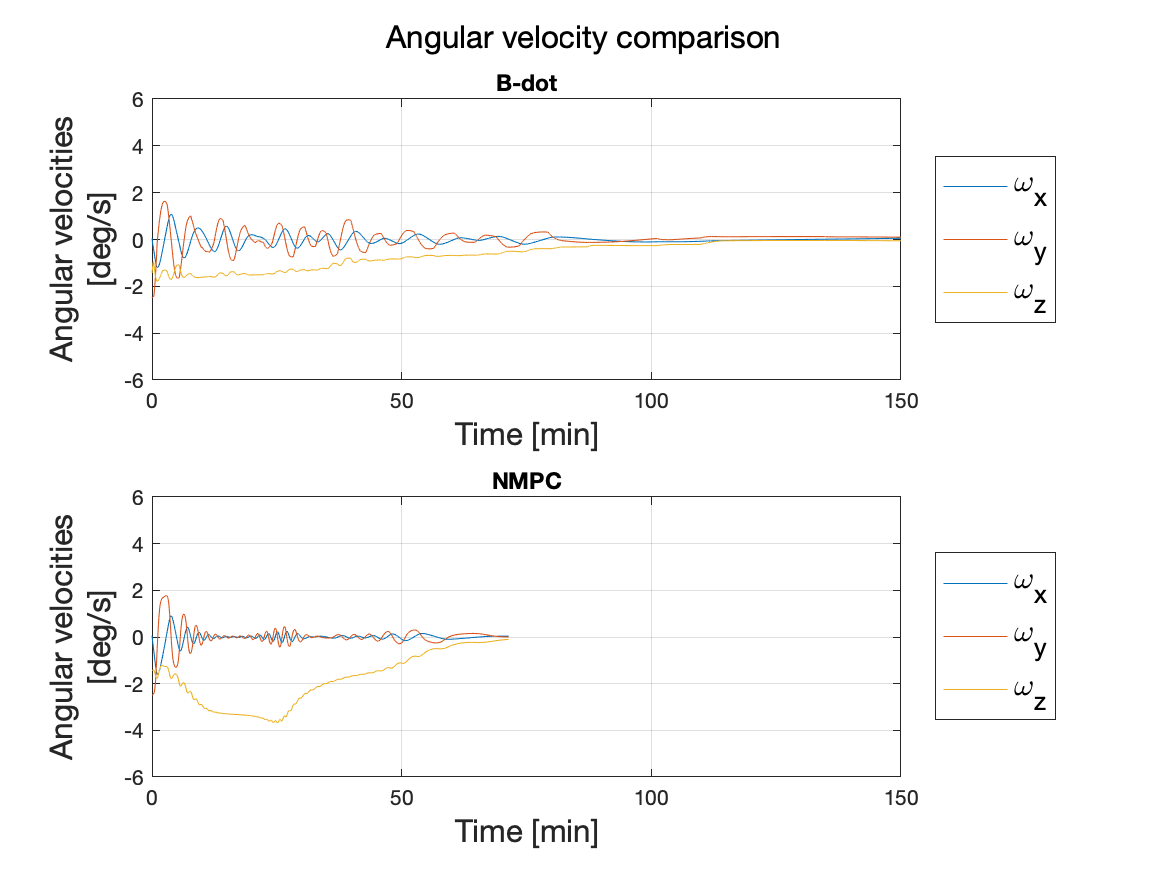}
\caption{Case 3: time history of angular velocities.}
\label{fig:Case3}
\end{figure}

\begin{figure}[H]
\centering
\includegraphics[width=0.75\textwidth]{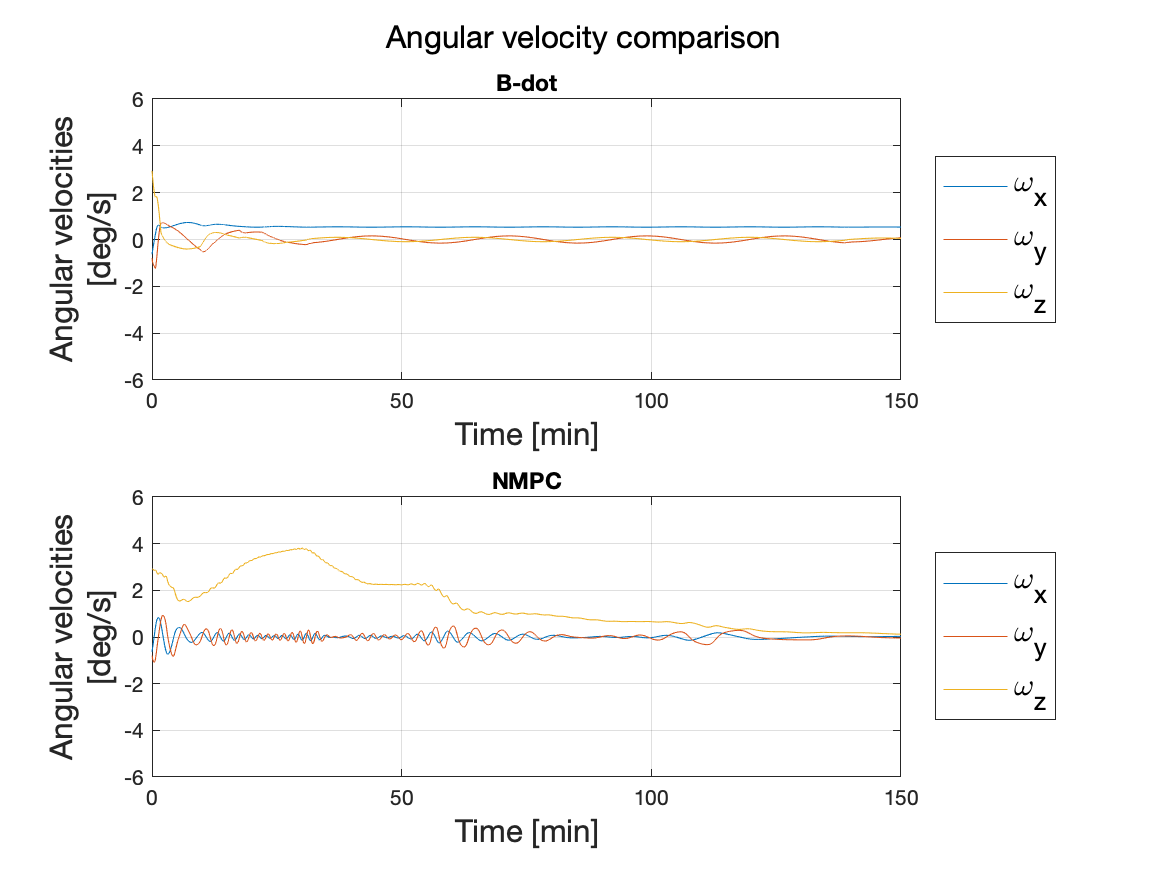}
\caption{Case 4: time history of angular velocities.}
\label{fig:Case4}
\end{figure}

Figure ~\ref{fig:Case1} to \ref{fig:Case4} present the simulation results from the four case studies. When the magnitude of all the angular velocities are less than 0.10 [deg/s], the simulations are set to be terminated in both B-dot and NMPC simulations. The simulations are run for the maximum time of 150 [min].

As can be seen in Fig.~\ref{fig:Case1}, although $\omega_y$ and $\omega_z$ are well attenuated, the B-dot algorithm is not able to sufficiently reduce $\omega_x$. NMPC, in contrast, is able to achieve detumbling.

Figure~\ref{fig:Case2} also reports an example where the B-dot law does not achieve all axes detumbling, but the NMPC algorithm does. Note that in the beginning of the trajectory, NMPC increases $\omega_z$ to be able to reduce $\omega_x$, which is a challenging variable to control. 

Figure~\ref{fig:Case3} is a case where both the B-dot method and the NMPC approach achieve detumbling. This requires about 150 [min] for the B-dot algorithm, while NMPC is able to achieve this within a much shorter time period.   

Finally, Fig.~\ref{fig:Case4} showcases a plot where neither the B-dot nor NMPC are able to detumble the spacecraft within the allocated time of 150 [min]. In the case of the B-dot algorithm, $\omega_x$ persists at a nonzero value. NMPC, on the other hand, is able to gradually attenuate all angular velocity components but is not able to fully detumble the spacecraft within the given
time.

\begin{figure}[H]
\centering
\includegraphics[width=0.75\textwidth]{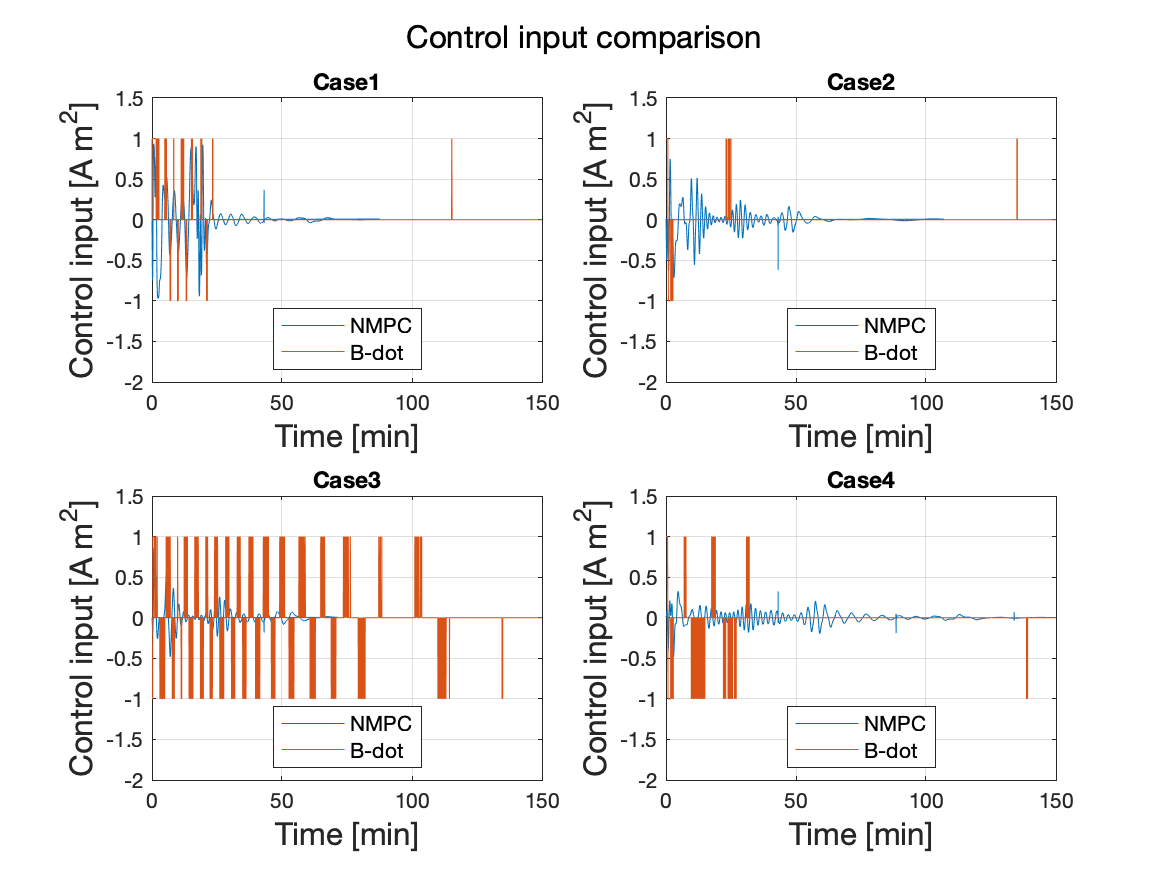}
\caption{Control input comparison.}
\label{fig:input}
\end{figure}

Figs.~\ref{fig:input} shows the time histories of control inputs in all the four cases. As can be seen, the NMPC controller requires much smaller control inputs. This can be advantageous in terms of reduced energy consumption and reduced electrical disturbance.

\section{Conclusions}

The satellite’s detumbling with only on a single-axis magnetic actuator is a challenging problem, in particular, requiring a different approach to stabilization than for three-axis and two-axis magnetic actuation systems. The necessary conditions for the controllability of spacecraft angular velocities involve: (1) spacecraft not having equal moments of inertia about the two principal axes that are orthogonal to the axis along which the magnetic actuator is aligned and (2) satisfying the specific rank controllability conditions derived in the note. The latter have been shown to hold numerically for the spacecraft and the orbit considered in the simulations. The classical B-dot law appears to be incapable of eliminating spacecraft rotational motion in the simulations, which has also been predicted from the theoretical analysis. The Nonlinear Model Predictive Control (NMPC) strategy based on the continuation/generalized minimal residual (C/GMRES) method has been shown to achieve detumbling within the allocated time through simulations in most cases. The possibility of detumbling spacecraft with only the single-axis magnetic coil opens the possibility for small spacecraft missions with stringent cost and packaging constraints. 

\section*{Appendix}

Below are the controllability distributions in Algorithm \ref{controllability_distribution_algorithm}:
\begin{eqnarray}
    \label{controllability_distribution_0}
    \triangle_{0}
    &=&
    {\rm span}\{f_{1}\}\\
    &=&
    {\rm span}\{[0, -\frac{B_z}{J_{y}}, \frac{B_y}{J_{z}}]^{T}\} 
\end{eqnarray}
\begin{eqnarray}\label{controllability_distribution_1}
    \triangle_{1}
    &=&
    \triangle_{0} \oplus \langle \triangle_{0}, f_{0} \rangle \oplus [\triangle_{0}, f_{1}]\\
    &=& \label{g_definition}
    {\rm span}\{f_{1}\} \oplus 
    \begin{bmatrix}
        \frac{1}{J_x J_z}B_y \omega_y (J_y - J_z) - \frac{1}{J_x J_y}B_z \omega_z (J_y - J_z)\\
        \frac{1}{J_y} \frac{\partial B_z}{\partial t} - \frac{1}{J_y J_z}B_y \omega_x (J_x - J_z)\\
        -\frac{1}{J_z}\frac{\partial B_y}{\partial t} - \frac{1}{J_y J_z}B_z \omega_x (J_x - J_y)\\
    \end{bmatrix}\\ 
    &=&
    {\rm span}\{f_{1},g_{1}\}
\end{eqnarray}
where $g_{1}$ is the second term in Eq.~\eqref{g_definition}.
\begin{eqnarray}\label{controllability_distribution_2}
    \triangle_{2}
    &=&
    \triangle_{1} \oplus \langle \triangle_{1}, f_{0} \rangle \oplus [\triangle_{1}, f_{1}]\\
    &=&
    {\rm span}\{f_{1},g_{1}\}
    \oplus {\rm span}\{ \langle f_{1}, f_{0} \rangle, \langle g_{1}, f_{0} \rangle \}
    \oplus {\rm span}\{ [ f_{1}, f_{1} ], [ g_{1}, f_{1} ] \}\\
    &=&
    {\rm span}\{f_{1}, g_{1}\}
    \oplus {\rm span}\{ g_{1}, g_{2} \} 
    \oplus {\rm span}\{ \boldsymbol{0}, g_{3} \}\\
    &=&
    {\rm span}\{f_{1}, g_{1}, g_{2}, g_{3} \}
\end{eqnarray}
where $g_{2}=\langle g_{1}, f_{0} \rangle$ and $g_{3}=[ g_{1}, f_{1} ]$. 
The explicit calculations of $g_2$ and $g_3$ give:
\begin{eqnarray}
    g_2&=&
    \begin{bmatrix*}[l]
         -\frac{2}{J_x J_y J_z}(J_y - J_z) (J_y \frac{\partial B_y}{\partial t} \omega_y - J_z \frac{\partial B_z}{\partial t} \omega_z)\\
        
        -\frac{1}{J_x J_y^2 J_z} (J_y (B_z J_x^2 \omega_x^2 - 2 \frac{\partial B_y}{\partial t} J_x^2 \omega_x - B_z J_x J_z \omega_x^2 + 2 \frac{\partial B_y}{\partial t} J_x J_z \omega_x - B_z J_x J_z \omega_z^2\\
        \ \ \ \ \ \ + \frac{\partial^2 B_z}{\partial t^2} J_x J_z + B_z J_z^2 \omega_z^2) - B_z J_x^3 \omega_x^2 - B_z J_z^3 \omega_z^2 + B_z J_x^2 J_z \omega_x^2 + B_z J_x J_z^2 \omega_z^2) \\ 
        
        \frac{1}{J_x J_y J_z^2} (J_z (B_y J_x^2 \omega_x^2 + 2 \frac{\partial B_z}{\partial t} J_x^2 \omega_x - B_y J_x J_y \omega_x^2 - 2 \frac{\partial B_z}{\partial t} J_x J_y \omega_x - B_y J_x J_y \omega_y^2\\
        \ \ \ \ \ \ + \frac{\partial^2 B_y}{\partial t^2} J_x J_y + B_y J_y^2 \omega_y^2) - B_y J_x^3 \omega_x^2 - B_y J_y^3 \omega_y^2 + B_y J_x^2 J_y \omega_x^2 + B_y J_x J_y^2 \omega_y^2)
    \end{bmatrix*}\\
    %
    %
\end{eqnarray}

\begin{equation}
    g_3=
    \begin{bmatrix*}
        \frac{2}{J_x J_y J_z}B_y B_z (J_y - J_z)\\
        0\\
        0\\
    \end{bmatrix*}
\end{equation}

\bibliography{bib.bib}

\end{document}